\documentclass[preprint2]{aastex}

\newcommand{\be}{\begin{equation}}
\newcommand{\ee}{\end{equation}}

\newcommand{\bee}{\begin{eqnarray}}
\newcommand{\eee}{\end{eqnarray}}

\usepackage{natbib}

\usepackage{color}

\usepackage{amssymb}
\usepackage[english]{babel}
\usepackage{ifpdf}
\usepackage[latin9]{inputenc}
\usepackage{mathpazo}
\usepackage{mathrsfs}
\usepackage{float}
\usepackage{psfrag}

\shorttitle{\textsc{HORIZON}}
\shortauthors{Zink, Kokkotas}

\begin{document}

\title{\textsc{HORIZON}: Accelerated General Relativistic Magnetohydrodynamics}

\author{B. Zink}
\affil{Theoretical Astrophysics, Institute for Astronomy and Astrophysics, Eberhard-Karls University of T\"ubingen, Germany}
\email{bzink@tat.uni-tuebingen.de}



\begin{abstract}
We present \textsc{Horizon}, a new graphics processing unit (GPU)-accelerated code to solve
the equations of general relativistic magnetohydrodynamics in a given spacetime.
We evaluate the code in several test cases, including magnetized Riemann problems and rapidly rotating neutron 
stars, and measure the performance benefits of the GPU acceleration in comparison to our
CPU-based code \textsc{Thor}. 
We find substantial performance gains in comparison to a quad-core CPU
both in single- and double-precision accuracy, and discuss these findings in the context of future
numerical modeling efforts.
\end{abstract}

\keywords{magnetohydrodynamics (MHD) --- methods: numerical}

\section{Introduction}

Computational models of general relativistic flows are an important tool to understand
the dynamics of compact objects, more so since gravitational wave detectors like
LIGO, VIRGO, TAMA or GEO600 are becoming operational. To facilitate interpretation
of data once they have been obtained, the inverse problem of determining
source parameters can be studied with the help of simulated models. 

General relativistic magnetohydrodynamics (GRMHD) is the study of magnetized flows
under relativistic velocities and in very strong gravitational fields, and it is therefore 
ideally suited for modeling compact objects and their environments (for a review see
\citet{Font:2008}). A number of GRMHD codes have been developed in recent years, 
see \citet{Koide:1999}, \citet{DeVilliers:2003}, \citet{Gammie03}, \citet{Komissarov04},
\citet{Anton05}, \citet{Mizuno:2006}, \citet{Anderson2006a}, \citet{Giacomazzo:2007ti},
\citet{DelZanna:2007}, \citet{Yamamoto:2008}, \citet{CerdaDuran:2008},
\citet{Bucciantini:2010} and \citet{Etienne:2010}. Most of these approaches employ finite-volume
methods on structured meshes, and use high-resolution shock capturing schemes
for handling the interface fluxes. Therefore, they are ideally suited for highly compressible
flows with shocks, and enjoy conservative properties which also ensure numerical stability.

Since these models are computationally expensive, not least due to their proper account 
of ultrarelativistic motions and the general covariant form of the conservation laws, 
there is a strong impetus to explore the use of graphics processing units (GPUs) to
accelerate the calculations. GPUs offer a substantially higher peak performance than 
traditional central processing units (CPUs) \citep{Kirk:2010}, and are better suited for the high throughput,
data-parallel problems typical of large-scale simulations. Therefore, these architectures 
have become very popular in recent years, especially since vendors like NVIDIA have 
started to target the high-performance computing market with specific hardware features 
and appropriate software tools. 

Graphics processing units have traditionally been designed for
computer graphics algorithms. Most of these algorithms are data-parallel,
i.e. the same operations are independently performed on a large stream of data. In contrast
to central processing units, which are optimized for executing single threads rapidly, GPUs,
as a data-parallel architecture, trade single-thread performance for massive parallelism. While
CPUs perform computations on a few cores, current-generation GPUs may utilise hundreds or even
over a thousand compute cores on a chip.

It is probably fair to say that GPUs do not only offer a much higher realizable performance 
at present, but also share architectural features which should become more prevalent in
\emph{any} hardware optimized for high throughput, data-parallel computations \citep{Garland:2010}. One reason
is given by power and cooling considerations. Traditionally, commodity CPUs have gained speed
by raising the frequency of operations, which gave rise to higher power consumption and more
challenging cooling requirements. This approach cannot be scaled indefinitely, and therefore
CPU designs have reached a \emph{power wall}. A second factor is the \emph{memory wall}: increasing
latencies in accessing main memory are covered in CPUs by using large caches and sophisticated
control logic, and the relative benefits of these approaches decrease with higher integration of the
circuits.

GPUs use massive parallelism in place of large caches and expensive control logic, hiding memory access
latencies with high-throughput computation on large streams \citep{Garland:2010}. Since more transistors
are devoted to compute units compared to CPUs, the peak performance
of GPUs (and similar architectures) is expected to increase substantially faster than in CPUs. 
Because of this, high-performance computing will greatly benefit from GPUs or similar designs in the
future.

Attempts to use GPUs for scientific applications have increased rapidly with the recent introduction 
of NVIDIA's CUDA language. Early adopters of this approach
in astrophysics include \citet{PortegiesZwart:2007}, \citet{Belleman:2008}, \citet{Zink2008short}, and 
\citet{Schive:2008}. \citet{PortegiesZwart:2007}, \citet{Belleman:2008} and \citet{Schive:2008} were concerned with
N-body calculations (also using GRAPE, another parallel architecture), whereas \citet{Zink2008short} performed
experiments on the general relativistic field equations. \citet{Barsdell:2010} and \citet{Fluke:2010} discuss
future uses of GPUs in astrophysical simulations, and \citet{Choudhary:2010} consider applications
to numerical relativity. \citet{Herrmann:2010} apply CUDA to the post-Newtonian equations describing
the binary-black hole problem, and \citet{Khanna:2010} investigate extreme-mass ratio inspirals. Specifically in the 
context of magnetohydrodynamics simulations, \citet{Pang:2010} compare the performance of
different hardware architectures. \citet{Wong:2009}, \citet{Wang:2009} and \citet{Wang:2010} present implementations 
of (Newtonian) MHD in CUDA, and observe very promising performance relative to CPU implementations.

In this paper, we present an implementation of general relativistic magnetohydrodynamics on graphics processing
units. The \textsc{Horizon} code is a standalone CUDA/C++ application which is in part based on the 
GRMHD code \textsc{Thor} \citep{Zink:2008b, Korobkin:2010}. In the following, we will first briefly describe the
physical system and the finite-volume method in section~\ref{sec:physical_system}, and then present the
\textsc{Horizon} code in section~\ref{sec:horizon_code}. In section~\ref{sec:performance}, we evaluate the
relative performance gain afforded by the GPU implementation compared to the CPU code. Afterwards, 
in section~\ref{sec:sp_vs_dp}, we illuminate how differences between single and double precision arithmetics
affect simulation results in a number of test cases. We conclude with a discussion in section~\ref{sec:discussion}.

\section{Physical system and numerical method}
\label{sec:physical_system}

\subsection{Evolution system}

The domain of \textsc{Horizon} are simulations of magnetized flows in general relativistic astrophysics. 
Similar to other codes in this field, e.g. \textsc{Thor} \citep{Zink:2008b}, \textsc{WhiskyMHD} \citep{Giacomazzo:2007ti},
\textsc{CoCoNuT} \citep{CerdaDuran:2008} and \textsc{Sacra} \citep{Yamamoto:2008}, this allows to investigate the properties and dynamics of ultra-compact sources
 of gravitational radiation, including
neutron stars, magnetars, and accretion flows around black holes.

In this paper we will only account for relativistic flows on a fixed spacetime
background, which is known as the Cowling approximation in the context
of stellar oscillations. Given such a spacetime metric $g_{\mu\nu}$, the conservation laws and Maxwell's 
equations are \citep{Font:2008}:
\bee
\nabla_\mu J^\mu & = & 0\\
\nabla_\mu T^{\mu \nu} & = & 0  \nonumber \\
\nabla_\mu F^{\mu \nu} & = & \mathcal{J}^\nu \nonumber \\
\nabla_\mu {}^* F^{\mu \nu} & = & 0 \nonumber
\eee
Here, $J^\mu = \rho u^\mu$ is the rest-mass density current, $T^{\mu \nu}$ is the 
energy-momentum tensor of the fluid and the Maxwell field, $\mathcal{J}^\nu$ is the 
electric four-current, $F^{\mu \nu}$ is the Faraday
tensor, and ${}^* F^{\mu \nu}$ is its Hodge dual.

We will assume the ideal MHD approximation to be valid in the following. Using a 3+1 split of Einstein's field equations, and an associated decomposition
of the GRMHD equations, we arrive at the conservation form \citep{Noble2006}
\bee
\label{eq:evolution_system}
\partial_t \left( \sqrt{-g} J^t \right) + \partial_i \left( \sqrt{-g} J^i \right)  = 0 \\
\partial_t \left( \sqrt{-g} {T^t}_\mu \right) + \partial_i \left( \sqrt{-g} {T^i}_\mu \right)
  = \sqrt{-g} {T^\kappa}_\lambda {\Gamma^\lambda}_{\mu \kappa} \nonumber \\
\partial_t \left( \sqrt{-g} B^j \right)  + \partial_i \left[ \sqrt{-g} (b^i u^j  -  b^j u^i) \right] = 0 \nonumber \\
\partial_i \left( \sqrt{-g} B^i \right) = 0. \nonumber
\eee
In these equations, we have used the magnetic field $B^i = {}^* F^{it}$, the Christoffel symbols ${\Gamma^\lambda}_{\mu \kappa} $,
and the magnetic four-vector $b^\mu$
is given in terms of the spacelike 3-hypersurface normal $n^\mu$ as $b^\mu = n_\kappa {}^* F^{\nu \kappa} 
({\delta^\mu}_\nu + u^\mu u_\nu) / n_\kappa u^\kappa$. We identify the eight conservative quantities 
$\sqrt{-g} J^t$, $\sqrt{-g} {T^t}_\mu$ and $\sqrt{-g} B^i$ which form the basis of the finite-volume 
method.

Initial data for the evolutions reported later will be either given
analytically (for Riemann problems) or will be imported from the publicly available \textsc{rns} code 
\citep{Stergioulas95} 
for the construction of rapidly rotating neutron star models. 

\subsection{Overview of the numerical method}
\label{sec:numerical_method}

We discretize the system of conservation laws eqns.~(\ref{eq:evolution_system}) with the use of 
a finite volume method and approximate Riemann fluxes at cell interfaces. This scheme has 
shock-capturing properties and is implemented in \textsc{Thor} and other GRMHD codes. 


The system of conservation laws (\ref{eq:evolution_system}) can be written in the compact form
\be
\partial_t w + \partial_i f^i(u(w)) = s(u(w))
\ee
where $w$ is the tuple of conserved variables ($\sqrt{-g} J^t$ and so on), $u$ is the tuple of primitive variables
($\rho$ and so on), and $f^i$ is the flux vector. For the finite volume approach, we discretize the coordinate domain into a regular mesh of
cells, and, for each cell, cast the conservation laws into a weak form by volume integration.

The evolution starts with converting the initial data, which is commonly given in terms of the primitive variables,
into the conserved variables. The relation $w(u)$ is algebraic and can therefore be easily evaluated in each cell.
For a time update, we need to evaluate the (time and area averaged) cell face flux vectors $f^i$. These are obtained
from assuming a local Riemann problem at the center location of the face, and calculating the flux
function appropriate for this problem. 

The initial data for the face Riemann problems are obtained from the cell center data via a 
reconstruction. To avoid unphysical oscillations in the reconstructions these approximants are constrained
by additional requirements (TVD and similar) which lead to extended reconstruction stencils compared to
Lagrangian interpolants of the same order. Each cell interface needs data reconstructed at its \emph{left} and 
\emph{right} side to define the local Riemann problem. 

After the initial state for the Riemann problem has been obtained, the appropriate flux is determined via a
local operator on each face. Since the full Riemann problem is rather expensive to solve (and even more so 
when magnetic fields are included \citep{Giacomazzo:2006}), we employ an approximate Riemann solver which
only requires the values of the flux function from each (left and right) reconstructed primitive state, and
the approximate maximal wave speeds associated with them.

When all face fluxes are known, we calculate the cell update from the effective flux per cell, and add the 
local source term contributions. We then have the new values of the conserved variables $w$, but
to be able to continue with the next operation we will also need consistent primitive variables $u$.
The inversion $u(w)$ is, unfortunately, not algebraic in general, so we employ a Newton-Raphson
method to obtain the primitive state. This concludes a (sub-)step of the Runge-Kutta
time advancement.

\section{The \textsc{Horizon} code}
\label{sec:horizon_code}

\subsection{Suitability of GPU architectures for general relativistic magnetohydrodynamics}

Since the computational operations in the finite-volume model described above are local (inside the support of a 
small stencil of only a few cells), the problem is naturally data-parallel via domain decomposition. This fact has been
used extensively in parallelizing GRMHD codes to distributed-memory architectures via MPI, such that each processor operates
on a compact subset of the full domain, e.g. a block of $40^3$ cells. Boundary information is then shared via synchronization
over \emph{ghost cells}, and every time new data is needed for reconstruction, MPI messages are exchanged between 
(geometrically) adjacent processes. Then, each process operates independently on its block using local loops.\footnote{This
method can potentially be extended by using OpenMP on a cluster node, thereby parallelizing intra-node operations
not via MPI but using the implicit communication employed by OpenMP. This approach can be employed to reduce the 
memory overhead involved with layers of ghost cells.}

This type of coarse-grained block decomposition, while data-parallel, is not sufficient for using graphics processing
units effectively. The number of processes in a typical MPI simulation will number in the tens or hundreds in many cases, whereas
GPUs require tens or hundreds \emph{of thousands} of threads to be saturated. The reason is that the simpler control logic and 
lower number of transistors devoted for advanced intermediate caches exhibit typically very high memory access latencies
(in the order of several hundred cycles), which are hidden in the data parallel model by oversubscribing hundreds or thousands
of compute cores with many more threads (see \citet{Garland:2010} for details).

Therefore we will employ a fine-grained method of parallelism: here, each compute core on the GPU operates either
on a single cell or a small set of cells. Given that a simulation may employ millions of cells in total, domain decomposition
performed in this way can easily provide enough threads to saturate a GPU. If several GPUs are used for a larger simulation,
a natural approach is to combine a coarse-grained decomposition into blocks as before, and then perform fine-grained
parallel computation on each GPU.

Synchronization on the level of the GPU, which in itself is a shared-memory architecture, is easy to perform by executing
independent compute kernels, and then using barrier statements to ensure that the global memory state on the device
is consistent. Between GPUs, MPI synchronization using ghost cells can in principle be used as before, with the added 
complication that current MPI libraries cannot directly access the GPU memory, such that the transfer of ghost cells from the
device memory to main memory (and back) must be performed before and after each MPI exchange.

Having addressed the matter of how to generate enough GPU threads and combine GPUs parallelism with MPI parallelism, 
a more challenging question concerns the kind of performance gains we can expect. Producing enough threads is easy
in a naturally data-parallel problem of large size, but this is not the only important consideration as
far as performance is concerned. 

The reason many threads are employed is to hide main memory access latencies, as mentioned above. However, to achieve
high performance the number of arithmetic (floating-point) operations in relation to main memory access operations,
the so-called \emph{arithmetic intensity}, must be high, so that the thread scheduler can run register-based operations
in certain threads while other threads are still waiting for data. This can be a problem for simple operations like 
matrix-vector multiplication (extensively used e.g. in some elliptic solvers) and even simple Newtonian fluid dynamics codes,
since the kernels then become limited by memory bandwidth, implying that processing cores are often idle waiting for data.

The theoretical memory bandwidth of a GPU is much higher than for a CPU, but it is only available
if the programmer is \emph{very} careful with data alignment and observes so-called \emph{coalescing rules} while reading from memory.
The latter refers to arranging memory operations within several threads in a particular way, such that load and store operations can
be performed in a faster manner. Ignoring the (at times rather arcane) rules of memory access coalescence can significantly
impact the available memory bandwidth (down to 1/10th in some architectures), and with it the available speedup
for any application which is bandwidth-limited.

These coalescing rules are one of the main reasons GPU programming is known to be challenging. In the field
of (Newtonian) computational fluid dynamics, which often has a low arithmetic intensity and, in addition, often needs to
operate on unstructured meshes, efficient compute kernels are a major challenge for software developers.
Some CFD methods, however, lend themselves to higher arithmetic intensities, e.g. Discontinuous Galerkin
techniques, and are therefore considered for GPU applications \citep{Kloeckner:2009}.

Fortunately, this is precisely where GRMHD is different. The relativistic equations of motion contain many more
operations per cell than is common for Newtonian finite-volume codes, even considering that more variables are needed
to store the spacetime metric. The higher arithmetic intensity of GRMHD reduces the relative pressure on memory
bandwidth from the compute kernels, and therefore can make use of a relatively larger portion of the (very high) compute
performance of the GPU \emph{without} excessive considerations of coalescing rules. The fact that recent 
architectures offer automatic caches makes the situation even more interesting for general-relativistic 
numerical codes.

There are additional performance considerations for writing fast GPU codes. Beside the memory coalescence, another
important factor concerns the so-called \emph{block size}. Current GPUs have a SIMD (single instruction multiple
data) architecture also on the hardware level, in the sense that a number of compute cores are arranged into 
multiprocessors with a common instruction decoder. In practice, the GPU operates on a set of threads (a \emph{warp})
in a single step, necessitating the exact same instruction to be performed up to the operands.
The programmer can still use conditional statements for convenience (in contrast to explicit vector instructions), 
but if the command unit detects a divergence within a warp, the number of different code paths are serialized, with
a comparable loss of available peak performance. For current architectures, this can be a factor of up to 32.

From the warp size, it is therefore desirable to have precisely common (essentially vectorized) execution parts at least
as far as multiples of 32 are concerned. This is a concern with standard GRMHD in as far as the recovery of primitive
variables are concerned, since the Newton-Raphson method may easily produce divergent execution paths depending
on the cell data. However, in all experience from CPU simulations we do not expect this part of the code to be important
for the overall speedup (even when considering Amdahl's law \citep{Amdahl:1967}), and in addition the loss of performance
from warp serialization competes with other possibly limiting factors. It is however important to arrange kernels
to execute threads in a multiple of the warp size to use all cores effectively.

Another factor affecting performance also has to do with the way threads are distributed onto different multiprocessors.
Each multiprocessor (remember that the GPU contains several such multiprocessors) contains fast local memory,
which provides cache storage, but also space for the registers used for floating point operations. If several blocks of threads 
are assigned to a multiprocessor, they have to share the common memory resources. In adverse cases, this 
could lead to \emph{register spilling}, where operands have to be ``cached'' into the device's main memory, thereby
strongly reducing available performance. Fortunately, it is possible to find a good thread configuration for each compute
kernel simply by a set of small experiments.

Even the shared memory accesses can have varying performance since the memory is organized into banks, and
random access to data can lead to bus conflicts and corresponding serialization. This, however, is often only a minor
consideration when compared to other factors affecting performance, and recent architectural developments for  
GPUs have further decreased its relevance.

We have, so far, left out one of the likely most important influences on GPU performance in current architectures: the
differences between single and double precision floating point operation cost. Historically, GPUs have evolved in the
context of graphics applications, which typically do not require more than single-precision accuracy even in demanding
situations. Very recent GPU architectures, which have been developed in recognition of the potential use of GPUs in scientific,
engineering and financial environments, have added support for double-precision operations, albeit at a higher cost in terms of 
clock cycles. In addition, bandwidth requirements for loading and storing double-precision numbers are naturally higher. The most
recent development in this direction is the Fermi architecture by NVIDIA which reduces the clock cycle difference
between single and double precision operations.


While the ratio between single and double precision performance may be reduced in future architectures, at present it is uncertain how
this difference will affect GRMHD codes in practice. It is quite possible that many operations in a scientific application do not actually
require double-precision accuracy, and only a select few operations, particularly those involving the (badly conditioned) subtraction
of similar but large numbers, should be cast to double precision floating point numbers. If the areas of code where this is necessary
can be identified, and if those are converted to double precision, the overall performance of the code may well be close to the one
for single precision, while the overall level of accuracy should be close to the double precision result. The effectiveness of such
a hybrid approach will depend on the numerical system. For GRMHD, we expect most operations to be well-conditioned, though
in particular the transformation of conserved to primitive variables may give rise to difficulties in some situations when using
single-precision floating point numbers (in fact, it could be argued that this statement also applies to double precision arithmetics).

Overall, we expect GRMHD to map very well to GPU architectures, due to its massively data-parallel nature, its mostly regular
memory access patterns (which facilitate caching), and its comparatively high arithmetic intensity resulting from the 
general relativistic set of equations.

\subsection{\textsc{Horizon}}

The \textsc{Horizon} code is an GPU implementation of general relativistic magnetohydrodynamics using the
discretization described in section~\ref{sec:numerical_method}. It supports two-dimensional and three-dimensional
meshes, and both single and double-precision floating point accuracy calculations. Particular kernels, in particular
the transformation from conserved to primitive variables, can be optionally performed exclusively in double precision,
even if the rest of the calculation is done with single precision numbers. The code employs HLLE (Harten, Lax, van Leer, 
Einfeldt) fluxes \citep{Harten83}, and a choice between TVD reconstruction with an MC limiter,
and PPM reconstruction (see \citet{Marti99} and \citet{Font:2008} for details). Primitive variables are recovered
using either the $1D_P$ or $2D_W$ methods from \citet{Noble2006}, and time is advanced via a third-order
Runge-Kutta method. A hyperbolic divergence cleaning scheme \citep{Anderson2006a} is available
for damping violations of the relativistic solenoidal constraint $\partial_i \left( \sqrt{-g} B^i \right) = 0$. 

A practical consideration when writing code for GPU architectures is the choice of parallel programming
language. Standard Fortran 90 or C can not be used for writing GPU code, but there are a number of 
language options available for data-parallel and stream programming problems. A very popular choice
is NVIDIA's CUDA for C, which is an extension of C/C++ to accommodate compute kernels
and functional programming on the device. In practice, CUDA uses standard C++ code enriched by
a number of additional keywords which identify functional parts to be executed on the GPU. Host (CPU) code
is extracted from the source and compiled with a native C++ compiler, e.g. GNU g++, whereas device
(GPU) code is transformed into a special intermediate language (similar to assembly) which is transformed 
for execution by a runtime library. From the programmer's point of view, these steps are all transparent,
which makes CUDA a particularly easy choice in practice.

The disadvantage of using CUDA is that it is a proprietary language which is only compatible
with NVIDIA GPUs. An alternative is OpenCL, which is designed as an industry standard to
replace proprietary solutions, but at this stage it is less mature, less convenient to use, and not quite as
well supported by the hardware drivers. In addition, there are a number of readily available numerical and 
parallel processing libraries for CUDA which are not ported yet to OpenCL. Overall, developing scientific
code is much more straightforward with CUDA at this time, which also explains its high popularity
in a number of scientific computing fields. 

\textsc{Horizon} is written in CUDA, and therefore based on object-oriented programming via C++. In particular,
in contrast to \textsc{Thor}, it is not part of the \textsc{Cactus} infrastructure, but is a standalone application code. 
It employs a Cartesian, uniform-mesh finite-volume discretization of GRMHD, distributed into host-side code
for initialization, memory allocation and input/output, and CUDA compute kernels for the mathematical operations.

The compute operations for evolutions are entirely performed on the CUDA device. This must be stressed, since
a simple model for accelerating an algorithm could be to first copy data to the GPU, perform computations, copy
back to host memory, and repeat these steps. However, memory transfers between host memory and the GPU device
(which use the PCI-Express bus) are rather slow compared to the GPU kernel execution time and available memory
bandwidth, so constantly copying data would severely limit the possible level of acceleration. In fact, there have been
cases where \emph{only} the kernel execution times have been used to measure speedups. \textsc{Horizon} runs entirely
within the GPU, and therefore the speedup mentioned here are what a practitioner would actually receive
in terms of overall execution speed of the entire application (when no output is considered). This is sometimes
also called \emph{application speedup}.

\subsection{Particular implementation aspects}

In the following, we will mention a few particular choices we have made to obtain high overall performance 
in \textsc{Horizon}. Even though GRMHD is a very suitable system for GPU acceleration, care must still be taken
in order to achieve high performance gains: see also \citet{Zink2008short} for a more detailed exposition
in the context of the ADM equations.

\subsubsection{Memory layout and data alignment}

We have discussed the issue of coalesced memory access above. While GRMHD has a high arithmetic intensity, it is
still very important to be aware of the major performance considerations concerning coalesced memory access
when writing compute kernels. While we will not be able to discuss the individual choices made for each kernel
here, we would like to point to two optimizations which are rather general and should apply directly
to similar implementations.

The first choice is to approximately satisfy the alignment requirements for coalesced memory access by
using properly ``padded'' memory blocks for three-dimensional meshes. CUDA supplies special functionality
to automatically align memory blocks in an appropriate way, which requires also a slightly different addressing pattern
inside the kernels. In cases where the grid size is not a multiple of the 
pad size defined by the device's coalescing rules, this will lead to a slight memory overhead, but also to increased
performance.

The second general consideration is to use blocks of threads which are properly aligned to allow coalesced accesses
to the padded grid function arrays. The SIMD nature of a multiprocessor already prescribes the threads per block to
be a multiple of the warp size to avoid warp divergence, but in addition coalesced access needs to read \emph{continuous}
blocks of padded data words using subsequent threads. Therefore, the access pattern employs subsequent threads
in the x direction by a multiple of the warp size, since the data layout is consecutive there. This also facilitates more
effective caching. Each block of threads then
operates on a slab of cells in x and y direction, e.g. (32,2) or (32,4), and performs a loop over the z direction.

\subsubsection{Use of automatic caches}

In older GPU architectures, no automatic caches were available, and therefore a set of memory shared between
threads inside a multiprocessor could be used to provide local data. The advantage is that (fast) coalesced memory
loads can be employed to stage this cached data, and then operate almost randomly on the shared data inside
the compute kernel. 

Effectively, the shared memory acted as a cache, but the programmer needed to stage and remove data manually, with
sometimes substantial (and often not very obvious) impact on the code's performance. This manual caching, while
still available, has been enhanced by the option to use automatic caches in NVIDIA's Fermi architecture. 
The programmer then has the option to devote more shared memory either to manually controlled storage, or rather to
the automatic cache, depending on the application problem and programming style.

\textsc{Horizon} exclusively uses the automatic cache when it is available on the GPU. This has simplified the implementation
considerably, and leads to very good performance results. While it is possible that manually optimized kernels using
specifically staged shared data could outperform our current implementation, we consider the effort not worth the possible
effect.

\subsubsection{Flux calculation and local arrays}

The flux calculation is very similar in the x, y, and z direction. Because of this, one may consider to use temporary
arrays (and code) for flux calculation in only one direction, but stage data by a map from 3D arrays into the
temporary arrays (which have a different data layout). Alternatively, a straightforward direction-dependent flux
calculation, which has a very different memory access pattern, can be used.

The advantage of one method over the other is not obvious: In the former case, compared to the latter, 
more data needs to shifted around, but the resulting (expensive) flux calculations can be done in a manner which 
is more consistent with the rules for coalescence. We have experimented with both methods, but found the second
approach to be superior on current Fermi architectures.


\subsubsection{Global reductions}

A practical implementation of a GRMHD code requires at some stage to perform different kinds of global operations,
e.g. to approximate integrals and build different norms. Many of these operations could be done in post-processing
on the CPU, but we have opted to also support global reductions on the GPU side. This is obviously more complicated
than in a serial code, since a parallel reduction can not be performed via a simple loop due to data dependencies.
For clusters, MPI libraries provide ready-made parallel reduction operations, but CUDA or other GPU languages do
not. The programmer can decide to construct a simple hierarchical reduction scheme manually, but it is easier
to rely on research on very efficient implementations of reductions, parallel scan and similar operations in the form
of parallel libraries. For CUDA, both the \textsc{CUDPP} and the \textsc{thrust} library are available.
In \textsc{Horizon} we employ the \textsc{thrust} library, a C++ template implementation
which offers a number of additional implementation benefits.

\subsubsection{MPI parallelization}

In order to make use of clusters of GPUs, we have added MPI support to \textsc{Horizon}. The approach is a standard ghost-zone
synchronization method which is also employed in \textsc{Cactus} and \textsc{Thor}, where we extend the computational
domain by a certain number of cells as appropriate for the numerical stencil of our scheme. The global domain
is divided into smaller blocks of comparable size, and each GPU in the cluster is assigned to an individual block.
In the synchronization step, we first prepare device-side buffers to hold the internal mesh information close to 
the boundaries (i.e. the information which needs to be transferred to the ghost cells of an adjacent domain block),
and invoke a kernel to copy mesh data to these buffers on each GPU. Then, the buffers, which are arranged in a linear
fashion inside the GPU's memory, are copied to the host memory. Afterwards, MPI messaging is used to exchange buffer
contents between adjacent blocks, and the buffer contents are copied back to the GPU memory. The final step invokes
another kernel to distribute this buffer data to the ghost cells on the local mesh.

\section{Performance results}
\label{sec:performance}

Our focus in this section will be on the performance differences between the CPU-based \textsc{Thor} code and the 
GPU-based \textsc{Horizon} implementation described in this paper. Our intent is to give practical insight into what can be gained from porting a GRMHD code to GPUs: it should
be clear that the particular numbers stated here are a consequence of (a) the particular hardware used, (b) our particular implementation
of the CPU-based \textsc{Thor} code, and (c) our particular implementation of the GPU-based \textsc{Horizon} code. Specifically,
it is possible that the CPU-based code, which is based on Fortran 90, could be further optimized. The results should therefore
be considered a guide to order-of-magnitude performance gains a practitioner can expect from using GPUs for
relativistic flow problems.

Nevertheless, we have endeavored to perform a fair comparison between the CPU and GPU codes by considering a number of
different factors: Firstly, the \textsc{Thor} code has more functionality than the \textsc{Horizon} code in that it supports multiblock meshes.
In particular, this requires the evaluation of local Jacobians and their derivatives in several places. For purposes of the comparisons,
we have deactivated this functionality entirely. 

Secondly, \textsc{Thor} is part of the \textsc{Cactus} computational infrastructure, while \textsc{Horizon} is
a standalone C++ application. Since \textsc{Cactus} is a powerful and general environment for hosting a variety of very different
modules with potentially complex interactions, it is quite possible that, even when we primarily use only one such module, 
the general computational overhead due to the infrastructure skews the comparison. We therefore normalize all numbers
obtained from \textsc{Thor} by comparing to an \emph{empty} evolution using \textsc{Cactus} in the sense that all computations involving
\textsc{Thor} are switched off.\footnote{In fact, this skews the numbers in disfavor of \textsc{Horizon}, since the calculations for
the time update using the method of lines are not regarded in this way, but they will be counted in the \textsc{Horizon} result.}
In practice, we have found this overhead to be small. 

Thirdly, we disregard time spent in the initial data setup for each code in the comparisons, and rather compare the wall clock
times spend on the main evolution loop. This is sensible since benchmark comparisons are made for very few iterations, 
whereas actual simulations run much longer, and therefore quickly amortize initial setup cost in typical cases. This is
certainly true for setup cost when we use shock tubes and \textsc{rns} data. A comparison which is predictive for long production
simulations should therefore exclude the setup costs.

Most tests here did not involve time spent in writing data to disk, i.e. we have switched off output. The reason
is that output is entirely limited by the (slow) disk access time, particularly when writing large quantities
of data, and this factor can be very different between hardware implementation and particular simulation parameters.
It is clear that a simulation which constantly writes a very large amount of data to a disk, e.g. to produce an animation,
will find that transfer a bottleneck for performance, and by Amdahl's law one would see little difference between CPU
and GPU implementations. However, in more typical situations, output is restricted to norms and integral quantities
every few time steps, and those should not have a large performance impact, provided the fast parallel reductions
discussed above are used on the GPU. We will present one test which estimates the performance impact of disk output
in a typical application situation.

The usual approach to measure relative GPU to CPU performance is by comparing a serial code running on the
CPU with a parallel code running on the GPU, resulting in the so-called \emph{parallel speedup}. This is a standard specification
of scaling behavior used in parallel environments, and it is simply defined as the ratio of the wall-clock time spent 
on the serial simulation with the wall-clock time in the parallel simulation. Since this is the number which
is being used in almost all literature on GPU performance, we will also report speedup measurements in this sense here.

However, modern CPUs are of course also mildly parallel, using typically four cores in present implementations, 
and therefore a more fair comparison could be to compare the speedup of the GPU-parallelized code with
a \emph{CPU-parallelized} code running on four cores. In fact, we think this is a comparison which should be
more relevant in practice, since modern 1U-sized cluster nodes usually have either four (quad-core) CPUs or 
four GPUs, and therefore such a comparison could reflect actual wall-clock time differences in practice more
faithfully. Therefore, we will also report these performance results in the following.

The CPU simulations have been performed on an Intel Xeon E5620 inside a dual-socket workstation. GPU architectures
used were an NVIDIA Geforce GTX 580, and a dual NVIDIA Tesla C2070 workstation for MPI tests.

\subsection{Timing results}
\label{sec:timing}

While performance comparisons could be done with almost arbitrary data, we will consider a setup which
is close to what we use in practice: the simulation of a rapidly rotating neutron star model. This is used e.g. 
for neutron star asteroseismology \citep{Zink:2010}, and as such the results stated here should be indicative
of real-world performance of GPUs in actual scientific applications. In this section we will focus on relative 
performance, while we consider the different accuracy of single and double precision calculations in section~\ref{sec:sp_vs_dp}.

The initial data is generated using the \textsc{rns} code \citep{Stergioulas95}: it is a standard uniformly rotating neutron star model
with a polytropic stratification (model BU2 from \citet{Dimmelmeier06a}). We will not discuss the details of this model since
they are not important for purposes of a performance analysis, but only state that it is based on a polytropic index $N = 1$,
has a mass of $1.46 \, M_\odot$ and a rotational period of about $2 \, \mathrm{ms}$.

We perform 100 iterations of the simulation at grid resolutions of $60 \times 60 \times 60$, $90 \times 90 \times 90$,
and $120 \times 120 \times 120$. We do not perform output to disk, and we normalize for initial data setup and 
computational overhead (the latter only in the CPU case).

\begin{table}
\begin{center}
\begin{tabular}{|l|r|r|r|}
\hline
   & \multicolumn{3}{|c|}{Grid size} \\
   & $60^3$ & $90^3$ & $120^3$ \\
\hline
CPU (1 core) & 372.9 s & 1278.2 s & 3110.8 s \\
\hline
CPU (4 cores) & 101.3 s & 326.4 s & 785.2 s \\
\hline
GPU (double pr.) & 9.1 s & 30.3 s & 60.1 s \\
\hline
GPU (single pr.) & 2.1 s & 6.7 s & 12.0 s \\
\hline
\end{tabular}
\end{center}
\caption{Rapidly rotating neutron star model: performance comparison 
between the CPU-based \textsc{Thor} code and the GPU-based \textsc{Horizon} code
for different problem sizes. All times are wall-clock times for 100 evolution steps.}
\label{table:comparison_rns_runtime}
\end{table}

\begin{table}
\begin{center}
\begin{tabular}{|l|r|r|r|}
\hline
   & \multicolumn{3}{|c|}{Grid size} \\
   & $60^3$ & $90^3$ & $120^3$ \\
\hline
GPU (SP) over & 177.7 x & 191.6 x & 259.2 x \\
CPU (1 core)  & & & \\
\hline
GPU (DP) over & 41.0 x & 42.2 x & 51.7 x \\
CPU (1 core) & & & \\
\hline
GPU (SP) over & 48.2 x & 48.9 x & 65.4 x \\
CPU (4 cores)  & & & \\
\hline
GPU (DP) over & 11.1 x & 10.8 x & 13.1 x \\
 CPU (4 cores) & & & \\
\hline
\end{tabular}
\end{center}
\caption{Relative performance (speedup) derived from table \ref{table:comparison_rns_runtime}.
The first row is the speedup measure usually employed in most GPU publications. However,
the third and fourth rows, each comparing either a GPU simulation in single precision (SP) 
or double precision (DP) floating point accuracy to a quad-core CPU simulation, should be more
relevant for a practical assessment of GPU over CPU performance.}
\label{table:comparison_rns_speedup}
\end{table}

Table~\ref{table:comparison_rns_runtime} shows the wall-clock time needed to perform this problem
on either one CPU core, four CPU cores (both double precision accuracy), on the GPU in double precision, and on
the GPU in single precision accuracy. Overall, the GPU offers a substantial increase in performance over both
single- and quad-core evolutions performed on the CPU.

The associated speedup factors are reported in table~\ref{table:comparison_rns_speedup}. It is evident
that the problem size has a substantial influence on the relative performance, an observation we have already made
in the case of an Einstein code \citep{Zink2008short}. It can be expected that a GPU, which is architecturally
based on running as many concurrent threads as possible to hide memory access latency, will perform
better on more threads (i.e. larger problem sizes). Nevertheless, part of the increased ratio could also
be attributed to an \emph{adverse} scaling of the CPU code with the problem size.

\begin{figure}[h]
\includegraphics[width=\columnwidth]{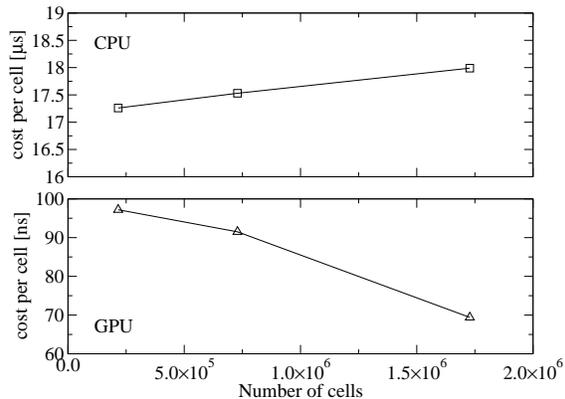}
\caption{Cost per cell update for different problem sizes. The upper panel shows the
CPU cost in microseconds, while the lower panel shows GPU cost (single precision)
in nanoseconds. The CPU performance is almost independent of the problem size,
while the GPU performs better with larger grids.}
\label{fig:cost_vs_grid_size}
\end{figure}

To investigate this, we calculate the computational cost per cell and iteration, and report the results in
figure~\ref{fig:cost_vs_grid_size}. A simple assumption of a machine's performance scaling
(that is, when we assume that only operation count is relevant for the cost) would imply
that is number is almost constant when varying the grid size. From the plot, it is apparent that the 
CPU code almost has this property. However, the GPU becomes more efficient with larger problem
size (i.e. a larger number of threads). It is therefore advisable to saturate the parallel processor
as much as possible to obtain improved overall performance.

\subsection{Relevance of disk output cost}

\begin{table}
\begin{center}
\begin{tabular}{|l|r|r|}
\hline
 & GPU (SP) & GPU (DP) \\
\hline
No output & 132.50 s & 635.54 s \\
\hline
With output & 134.06 s & 637.46 s \\
\hline
Relative cost & 1.17 \% & 0.30 \% \\
\hline
\end{tabular}
\end{center}
\caption{Impact of disk output on GPU simulation performance, using an output
frequency typical of a simulation performed in neutron star asteroseismology. 
GPU simulations have been performed in both single (SP) and double precision (DP),
with the expectation that the relative cost is lower in double precision.}
\label{table:disk_output}
\end{table}

In this section, we consider the impact of disk output on the performance result. As discussed before, writing
arbitrarily large quantities of data to disk would make the disk access the bottleneck in the problem,
and in such a case neither CPU nor GPU speed may be very relevant. In contrast, we select a set of
output parameters typical for our own scientific simulations \citep{Zink:2010} and observe the 
performance impact of the output. For this test, we evolve a neutron star for a time interval
of $1 \, \mathrm{ms}$ with a time step of $0.5 \, \mu\mathrm{s}$. Output of overall norms (maximal
value of density and similar) is performed every $5 \, \mu \mathrm{s}$, whereas one-dimensional
profile output along each axis, and also three-dimensional output of most evolution quantities, is
performed every $50 \, \mu\mathrm{s}$. The test is performed on a regular workstation with a SATA
disk: a dedicated high-performance RAID system is expected to exhibit a smaller impact on performance.
We report the results of the comparison in table~\ref{table:disk_output}. As can be seen, 
output does not substantially affect performance in this typical application case.

\subsection{Performance on multiple GPUs}

\textsc{Horizon} supports MPI parallelization, and therefore it is interesting to investigate the cost of scaling
to multiple GPUs. At the time of writing, we have access to a dual Tesla C2070 setup inside a single 
workstation, and the following observations will therefore compare performance between a single and two
GPUs.

We have tested the MPI scaling performance of \textsc{Horizon} using a grid size of $120 \times 120 \times 120$
for a single GPU, and $240 \times 120 \times 120$ for two GPUs. This choice comprises a weak scaling test, which
should be appropriate in the present context since it excludes the grid-size dependent performance of
the GPU (see figure~\ref{fig:cost_vs_grid_size}) from the consideration of MPI messaging overhead
measurements. The actual initial data is of little consequence in the measurement: because of the grid setup,
we have opted for a Riemann problem.

\begin{table}
\begin{center}
\begin{tabular}{|l|r|r|}
\hline
   & SP & DP \\
\hline
1 GPU & 17.74 s & 79.81 s \\
\hline
2 GPUs (MPI) & 18.38 s & 80.94 s \\
\hline
Parallel efficiency & 96.5 \% & 98.6 \% \\
\hline
\end{tabular}
\end{center}
\caption{Cost of MPI parallelization (parallel efficiency), comparing
a simulation running on two GPUs with a single GPU simulation,
both in single (SP) and double precision (DP). This is a nontrivial test
for GPUs even when only two MPI processes are used.}
\label{table:mpi_efficiency}
\end{table}

The results of the timing tests are reported in table~\ref{table:mpi_efficiency}. As for the
disk output case, the parallel synchronization is a serial point inside the parallel program,
and therefore could be a limiting factor by Amdahl's law. However, in contrast to disk output,
the MPI synchronization needs to be performed \emph{at every substep} of the method to
have a consistent global state. In principle, this could give rise to a substantial reduction
in overall performance, but in the present case we observe good scaling efficiency. These results,
while encouraging, are of course only partially indicative of parallel efficiency on a full cluster.
However, given a particular (large) problem size, substantially less GPUs are needed to obtain
the same performance as CPUs, thus requiring less MPI processes in comparison.

\section{Comparison between single and double precision accuracy}
\label{sec:sp_vs_dp}

This section will repeat selected standard test cases for GRMHD already performed with 
\textsc{Thor} using the GPU-accelerated \textsc{Horizon} code, with a focus on differences
between single and double precision floating point accuracy. This is of particular interest
as a consequence of the performance gap between these options, and whether a hybrid programming
model, i.e. one that uses single precision accuracy in most operations and double precision
in only selected ones, could be feasible. The results reported here use either pure single or double
precision accuracy operations.

\subsection{Shock tubes}

We will first consider the \emph{Sod test} \citep{Sod:1978}, which is a standard test case for compressible hydrodynamics codes. 
We prepare a Riemann problem in an ideal gas with $\Gamma = 5/3$, and with initial
states $\rho_L  =  1$, $P_L = 1$,  $u^i_L = 0$, and $\rho_R = 0.125$, $P_R = 0.1$, $u^i_R = 0$.  We prepare this
problem on a full $120 \times 120 \times 120$ grid using \textsc{Horizon}, and simulate the evolution up
to a time $t = 0.8$. For comparison, a reference solution using \textsc{Thor} has been produced using a very
fine grid.

\begin{figure}[h]
\includegraphics[width=\columnwidth]{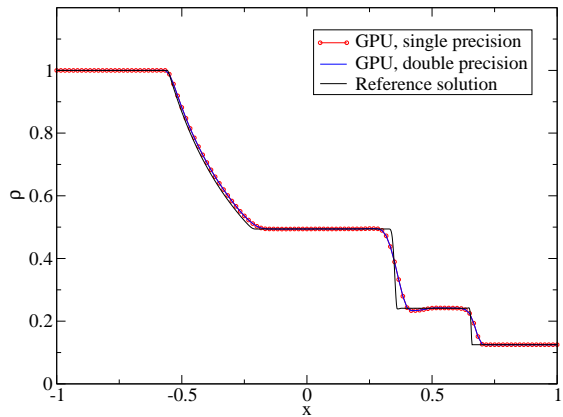}
\caption{Sod test: comparison between single and double precision accuracy. This plot shows the
density profile at time $t =0.8$. The GPU results were obtained with \textsc{Horizon}, whereas the
reference solution is produced with \textsc{Thor} on a very fine grid.}
\label{fig:sod_rho_x}
\end{figure}

\begin{figure}[h]
\includegraphics[width=\columnwidth]{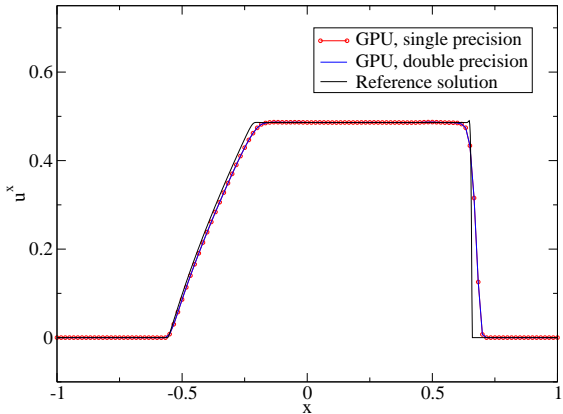}
\caption{Same as figure~\ref{fig:sod_rho_x}, but comparing the x velocity $\tilde{u}^x$.}
\label{fig:sod_ux_x}
\end{figure}

Figures~\ref{fig:sod_rho_x} and \ref{fig:sod_ux_x} show the density and x-velocity profiles at the end
of the evolution. The profiles show the somewhat dissipative nature of the HLL solver and TVD limiter
we are using, but in particular, they show no discernible difference between single and double precision
results. In fact, these differences are $|\delta \rho| \approx 5 \times 10^{-5}$ and 
$|\delta \tilde{u}^x| \approx 10^{-4}$ exactly at the location of the shock, and typically
$|\delta \rho| < 1 \times 10^{-6}$ and $|\delta \tilde{u}^x| < 10^{-6}$ in smooth parts of the flow.
These errors are at or lower than the level of the discretization error of the scheme.

A set of tests for magnetized flows have been proposed by \citet{Balsara:2001}. We will show results from
the first Balsara test, with initial data $\rho_L = 1$, $P_L = 1$, $u^i_L = 0$, $B^x_L = 0.5$, $B^y_L = 1$,
$B^z_L = 0$, and $\rho_R = 0.125$, $P_R = 0.1$, $u^i_R = 0$, $B^x_R = 0.5$, $B^y_R = -1$, $B^z_R = 0$.
The setup is otherwise the same as for the Sod test.

\begin{figure}[h]
\includegraphics[width=\columnwidth]{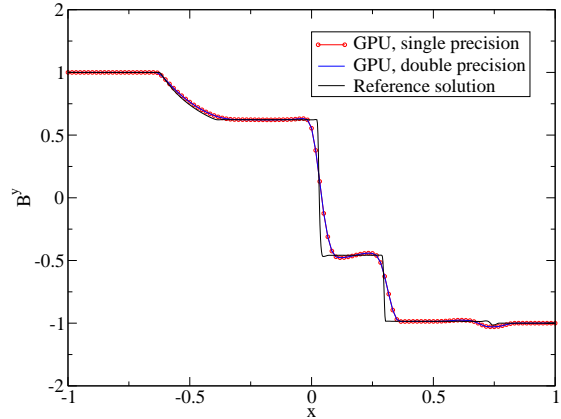}
\caption{Balsara 1 test: comparison between single and double precision accuracy. This plot shows the
magnetic field $B^y$ profile at time $t =0.8$. The GPU results were obtained with \textsc{Horizon}, whereas the
reference solution is produced with \textsc{Thor} on a very fine grid.}
\label{fig:balsara1_By_x}
\end{figure}

In figure~\ref{fig:balsara1_By_x} we show a comparison of the magnetic field vector component $B^y$
(which is initially discontinuous) at the end of the evolution, as a representative quantity for comparison. 
As in the case of the Sod test, also the magnetic field structure is not significantly affected by using single
precision accuracy, at least in this test case. The absolute differences between single and double precision
are $|\delta B^y| < 10^{-6}$, and similar statements hold for all other evolved quantities.

\subsection{Oscillations in a rapidly rotating neutron star}

A next comparison between single and double precision accuracy will be performed using a slightly perturbed
stellar model. We use the model BU2 as in section \ref{sec:timing}, but additionally perturb it
with a small quadrupole deformation ($\ell = 2, m = 0$) in order to analyze the dynamics of the star.

\begin{figure}[h]
\includegraphics[width=\columnwidth]{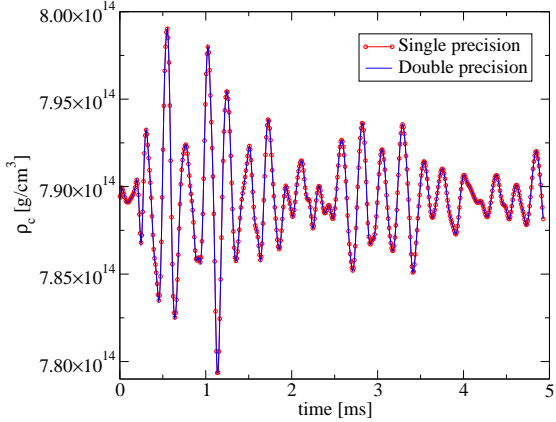}
\caption{Oscillations of a rapidly rotating neutron star model: comparison between single
and double precision evolutions using the \textsc{Horizon} code. This plot shows the central
density of the star over a time of $5 \, \mathrm{ms}$.}
\label{fig:rns_rho_maximum_sp_vs_dp}
\end{figure}

\begin{figure}[h]
\includegraphics[width=\columnwidth]{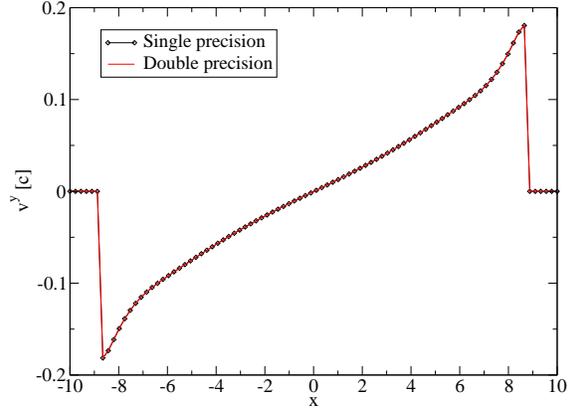}
\caption{Oscillations of a rapidly rotating neutron star model: comparison between single
and double precision evolutions using the \textsc{Horizon} code. This plot shows the evolved
rotational velocity profile $\tilde{u}^y (x)$ after an evolution of $5 \, \mathrm{ms}$.}
\label{fig:rns_vy_x_sp_vs_dp}
\end{figure}

The tests are performed with a moderate resolution ($90 \times 90 \times 90$, with about $60-70$ cells covering the
stellar diameter), and we evolve the star for $5 \, \mathrm{ms}$ physical time, corresponding to $10,000$ iterations. 
Figure~\ref{fig:rns_rho_maximum_sp_vs_dp} shows a comparison of the central density evolution using 
single and double precision simulations. As can be seen, the differences are very small, and likely smaller than the
discretization error of the system. When transforming both time profiles into Fourier space (not shown here), 
we similarly get virtually identical spectra, and can identify the fundamental radial mode $F$ at $2596 \pm 30 \, \mathrm{Hz}$,
its overtone $H_1$ at $4403 \pm 30 \, \mathrm{Hz}$, the fundamental quadrupole mode ${}^2f$ at $1909 \pm 30 \, \mathrm{Hz}$,
and its overtone ${}^2p_1$ at $3970 \pm 30 \, \mathrm{Hz}$. These numbers match very well with \citet{Gaertig08a}, but specifically, 
there is no discernible difference between the single and double precision results.

Another very sensitive quantity is the rotational velocity profile of the star. In figure~\ref{fig:rns_vy_x_sp_vs_dp}, we
compare the x-axis profile of $\tilde{u}^y (x)$\footnote{We use the primitive velocity variables of \citet{Noble2006},
in which the four-velocity $u^\mu$, the Lorentz factor $\gamma$, the lapse function $\alpha$ and the shift vector $\beta^i$ 
induce spatial velocities $\tilde{u}^i = u^i + \gamma \beta^i / \alpha$, which are related to the quantities $v^i$ \citep{Banyuls97} used 
in some other GRMHD codes by the Lorentz factor. However, this distinction should not be important
in the present context.} after $5 \, \mathrm{ms}$. Again, virtually no differences between single and double precision
results are apparent, even in the problematic region near the stellar surface. The maximal deviation is
$|\delta \tilde{u}^y| < 5 \times 10^{-6}$. 

We conclude that, at least for the applications presented here, single precision calculations should be sufficient.
However, as with all other parameters entering a numerical calculation (discretization scheme, time step, 
spatial resolution, \ldots), this assumption must be tested on a case-by-case basis. We will further discuss this
point below.

\subsection{Strongly magnetized neutron star}

\begin{figure}[h]
\includegraphics[width=\columnwidth]{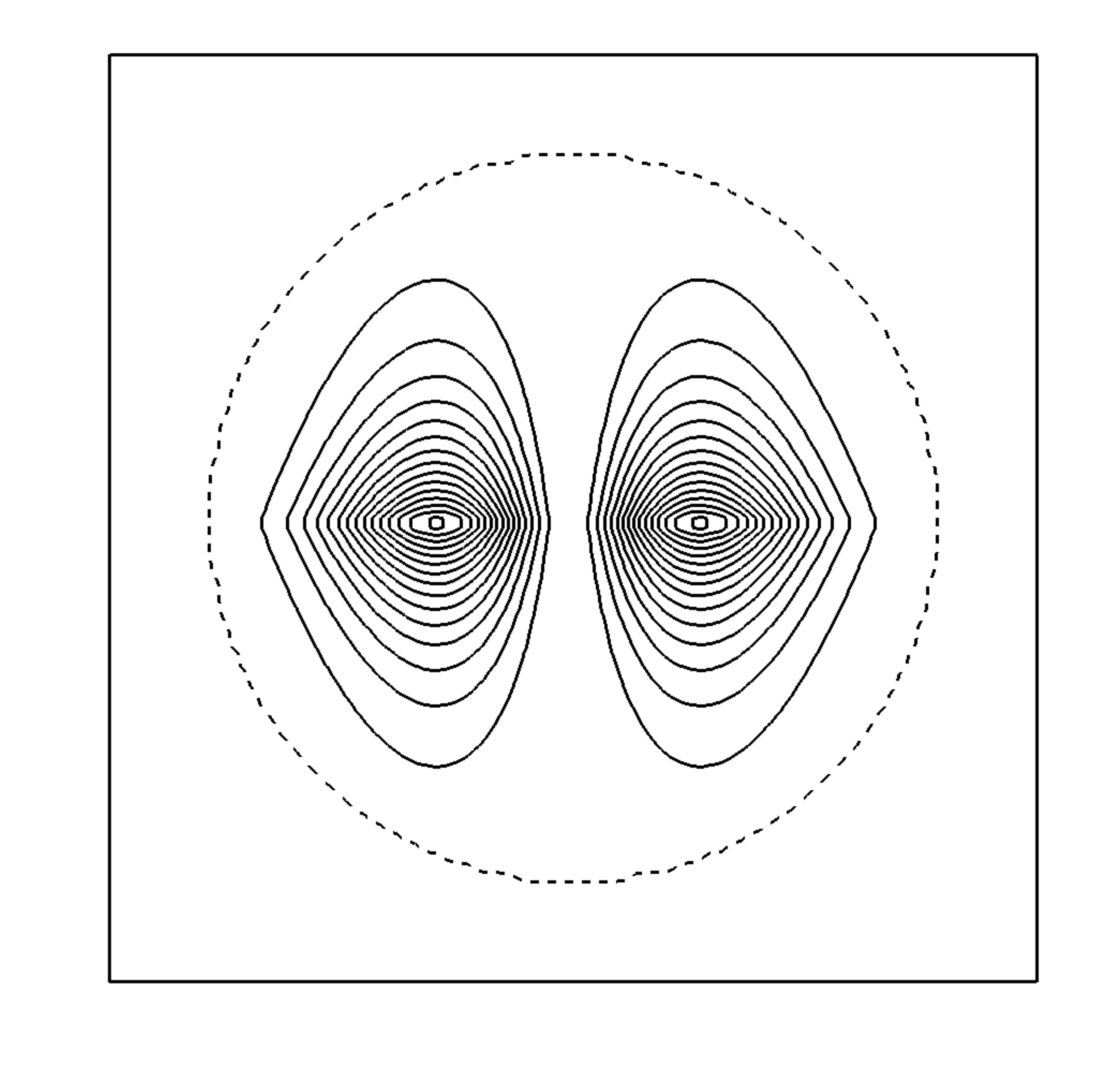}
\caption{Strongly magnetized neutron star: Initial distribution of the amplitude of the magnetic
field within a meridional cut, with contours up to about $2 \times 10^{15} \, \mathrm{G}$. The dashed line denotes the stellar
surface.}
\label{fig:magn_rns_initial_pol}
\end{figure}

\begin{figure}[h]
\includegraphics[width=\columnwidth]{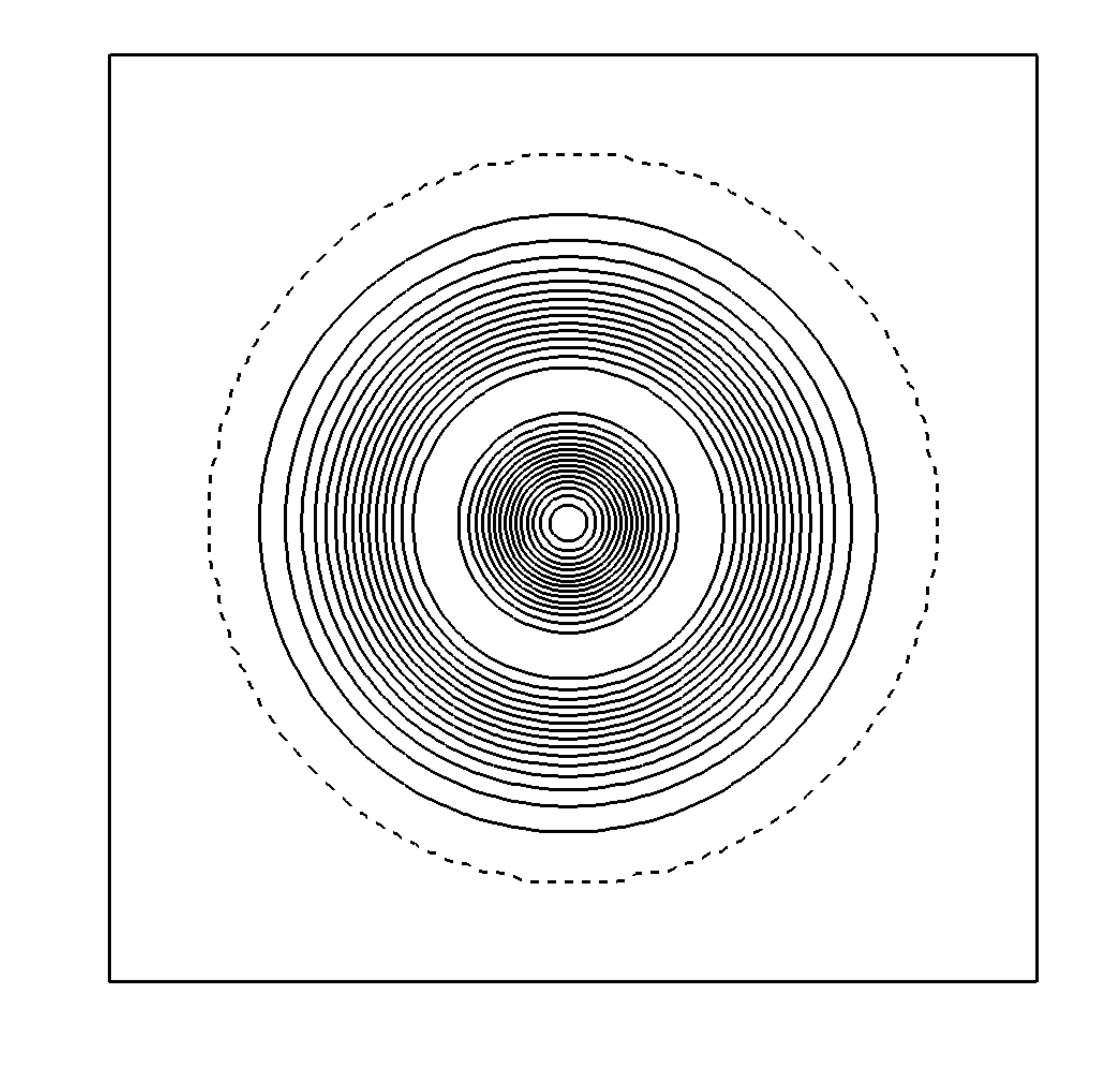}
\caption{Same as figure~\ref{fig:magn_rns_initial_pol}, but inside the equatorial plane of the star.}
\label{fig:magn_rns_initial_equ}
\end{figure}

As a final point of comparison we explore a magnetized neutron star model. For this test, we use the
nonrotating model BU0 \citep{Dimmelmeier06a}, which is defined by $\rho_c = 1.28e-3$ and $\Gamma = 2$
and has a mass of $M = 1.4 \, M_\odot$. On this model, we superimpose a toroidal magnetic field given 
by \citep{Braithwaite:2006}
\be
\mathbf{B} = B_0 \left(\frac{\varpi}{\varpi_0} \right)^2 e^{-\varpi^2/\varpi_0^2} e^{-z/2H} \mathbf{e}_\phi
\ee
where $\varpi = \sqrt{x^2 + y^2}$, and we choose $\varpi_0 = 3$, $H = 1$, and $B_0$ such that the maximal field strength is 
approximately $2.6 \times 10^{15} \, \mathrm{G}$. The structure of the initial magnetic field is shown in figures~\ref{fig:magn_rns_initial_pol}
and \ref{fig:magn_rns_initial_equ}.

Even with these comparatively high field strengths, the magnetic field is only
a small perturbation to the fluid equilibrium. It is also known that this field is unstable \citet{Braithwaite:2006, Lander:2010, Duez:2010},
but we will only be concerned here with differences between single and double precision evolutions. As in the previous
section, we evolve the model for $5 \, ms$ physical time on a $90 \times 90 \times 90$ grid.

\begin{figure}[h]
\includegraphics[width=\columnwidth]{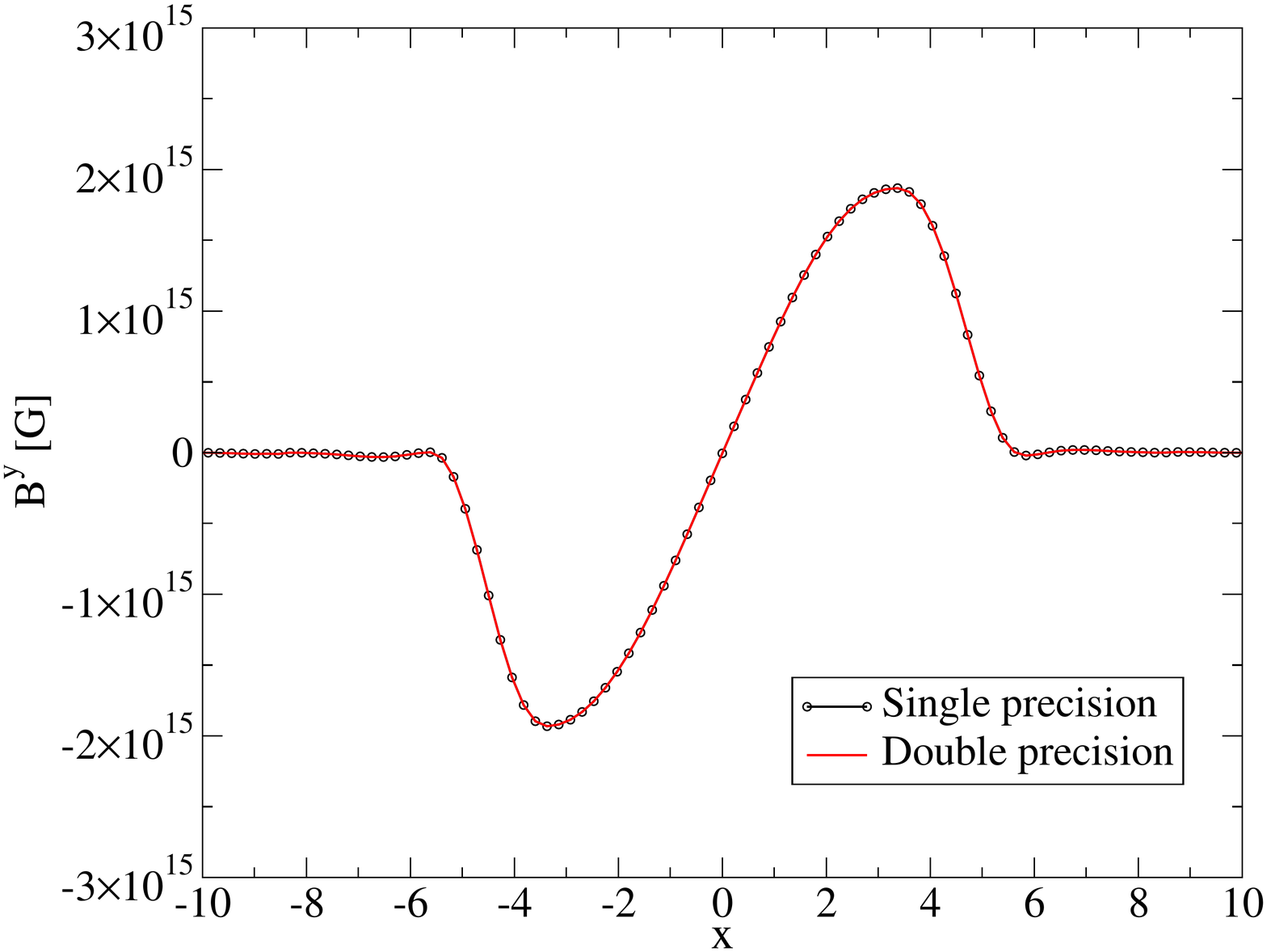}
\caption{Strongly magnetized neutron star: Comparison between single and double precision
evolutions. This plot shows the profile $B^y (x)$ after an evolution of $5 \, \mathrm{ms}$.}
\label{fig:magn_rns_By_x_sp_vs_dp}
\end{figure}

Figure~\ref{fig:magn_rns_By_x_sp_vs_dp} shows the profile $B^y (x)$ at
the end of the evolution from both simulations. The disagreement is within 
$|\delta B| < 10^{12} \, \mathrm{G}$, and therefore larger than in the non-magnetized case,
but still well below $10^{-3}$ of the actual field strength.

\section{Discussion}
\label{sec:discussion}

\begin{figure}[h]
\includegraphics[width=\columnwidth]{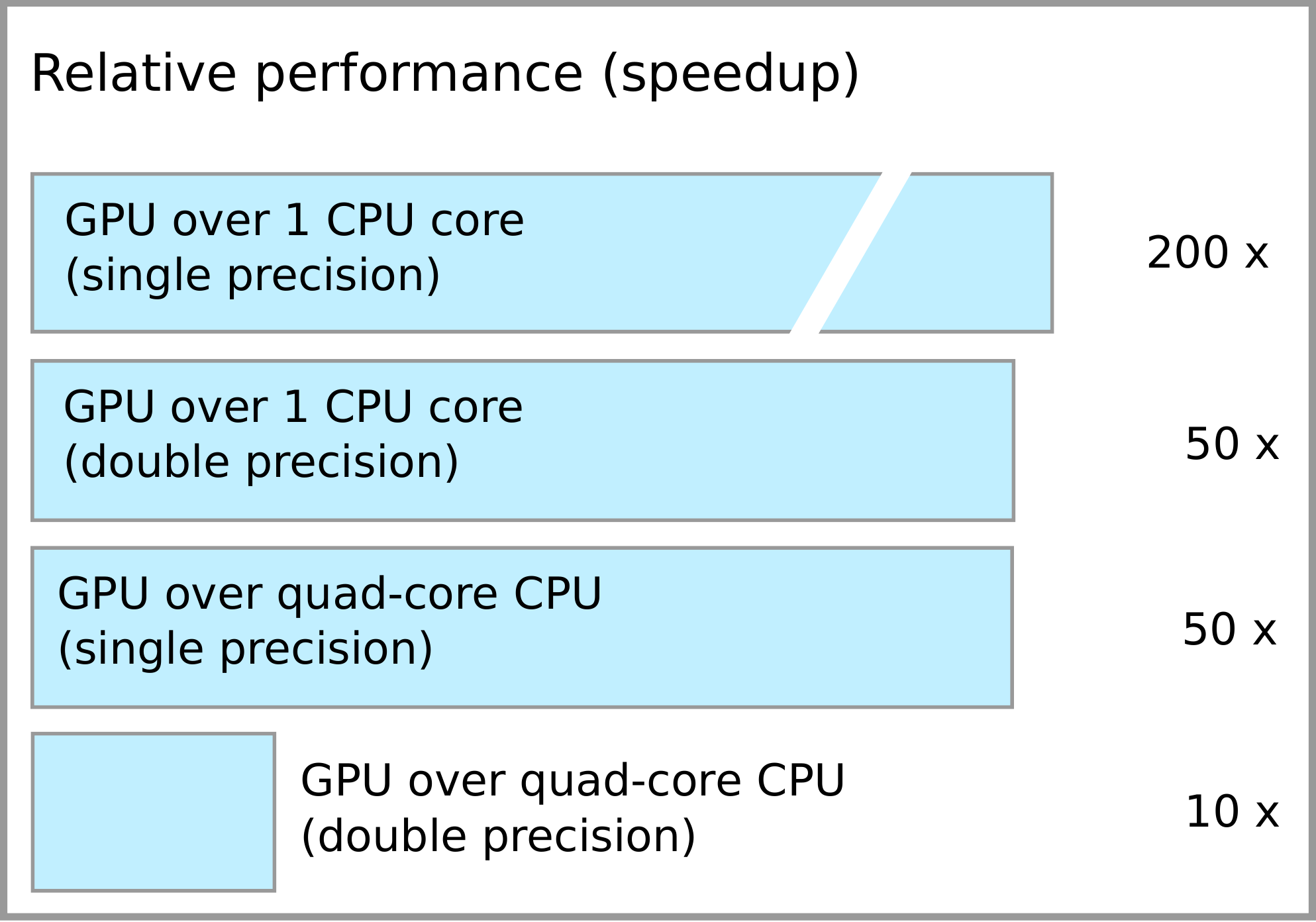}
\caption{Relative performance of the GPU-based \textsc{Horizon} code compared to the CPU-based
\textsc{Thor} code. These numbers approximately represent the timing results obtained in section~\ref{sec:performance},
which have been measured with an Intel Xeon E5620 CPU and an NVIDIA Geforce GTX580 GPU.
The first number is the one typically used in publications on GPU codes, whereas the third and fourth
row show more realistic comparisons between a parallel code running on a GPU and a parallel
code running on a quad-core CPU. In section~\ref{sec:sp_vs_dp} we have presented results indicating
that single precision accuracy may be sufficient for a number of applications. Further discussion of this
point can be found in the text.}
\label{fig:gpu_speedup}
\end{figure}
 
This paper describes an implementation of general relativistic magnetohydrodynamics on graphics processing units, 
the \textsc{Horizon} code, and presents results on the relative performance practitioners can expect to gain
from employing GPUs in such a context. The main results from performance measurements, obtained in a setup
which is close to applications in neutron star asteroseismology, are summarized in figure~\ref{fig:gpu_speedup}.

We have argued that general relativistic magnetohydrodynamics should map very well to GPUs as a consequence
of the data-parallel nature of the problem, mostly regular memory access patterns, and a higher arithmetic intensity than is 
common in traditional Newtonian MHD codes. Timing tests have confirmed this expectation, and have shown substantial
performance gains over current CPU architectures. We have selected a particular CPU/GPU combination for these tests,
but this general picture will not change when comparing different sets of current hardware implementations, even though
the actual speedup numbers will vary.

From an astrophysical application perspective, we can expect about an order of magnitude increase in performance by
employing a double precision GPU calculations in place of a quad-core CPU, and several orders of magnitude when employing
only single precision arithmetics. This is a direct consequence of the massively parallel, high-throughput nature of the GPU,
and the fact that GRMHD maps well to this kind of architecture.

Given the often very demanding problems which result from implementation of astrophysical models, and the high cost
of three-dimensional explicit simulations, an order of magnitude in performance should be enough of an incentive to
consider GPUs, and similar massively parallel architectures, in place of CPUs. Such a performance gain is interesting
to attack substantially larger simulations (or longer evolution times) on compute clusters, provided a GPU-based machine is
available. 

However, another use of GPUs should be mentioned which is no less important in practice: the ability to
perform reasonably sized full-scale simulations \emph{on a single workstation}, with a high turnaround time for
results. This is relevant to explore new numerical methods and model parameters spaces, which can now
be performed much faster, and much more conveniently, than using many nodes on a cluster resource. In this way, 
scientific productivity can be gained very easily with a modest investment in hardware.

The much higher throughput obtained in single precision computations begs the question whether computations
could also be performed with a reduced level of accuracy. We would first like to stress that this aspect of course also applies
to the distinction between double and quadruple precision, and there may well be problems for which even double precision is
not sufficient. Overall, this is entirely dependent on the particular problem and numerical algorithm employed, and 
only experimentation can establish whether a particular floating point approximation is sufficient. Of course, this equally
applies to all other free parameters of the discretization.

In the particular cases we have investigated here, which consider shock tubes, small oscillations of rotating stars,
and strongly magnetizes stars, it seems that single precision floating point accuracy could be sufficient for a number of modeling purposes. That statement may well cease to
be true if tabulated equations of state, turbulence, radiation transport, reaction networks etc. are involved in a calculation.
If such is the case, a good strategy could be to locate (approximately) the components of the numerical code which
should require double precision accuracy, and then employ the higher accuracy only in these parts. \textsc{Horizon} offers
exactly this option by allowing to adjust the accuracy of the conservative to primitive variable transformation and the calculation
of flux differences. Such a hybrid approach could preserve very high levels of performance while keeping the required level 
of accuracy.

But the most important aspect of GPU computation, and more generally, massively data-parallel hardware architectures (which may
also be located on a CPU), are not the potential gains which could be achieved at the present time: it is the fact that the peak performance
in these hardware designs \emph{grows much more rapidly} than the peak performance of traditional CPUs. That is, the relative factor of 10
we can achieve today may translate into a factor of 20 in two years time. Preparing for this major shift in high-performance architecture,
which is not only driven by the isolated requirements of a small market segment as in the past, but pervades the global development 
of all consumer and professional-level microprocessors, should be considered a priority for future computational tools
in astrophysics.

\acknowledgments{The author gratefully acknowledges discussions with and feedback from E. Gaertig,  F. Herrmann, 
K. Kokkotas, P. Lasky, E. Schnetter, N. Stergioulas, and M. Tiglio. This work was supported by the 
Sonderforschungsbereich/Transregio 7 on gravitational wave astronomy by the DFG.}

\bibliographystyle{apj}

\end{document}